\documentclass{Rinton-P10x7}
\usepackage{psfig}

\begin{document}

\title{Correlators in ${\cal N}=1$ SYM(2+1)
\thanks{Work with S.~Pinsky and J.~Hiller.}}

\author{Uwe Trittmann}

\address{Department of Physics, Ohio State University, 
174 W 18th Ave, Columbus, OH 43210, USA}

\maketitle

\abstracts{
We present a calculation of the correlator 
$\langle 0|T^{++}(r) T^{++}(0) |0\rangle $ in
${\cal N} =1 $ SYM theory in 2+1 dimensions. In the calculation, we 
use supersymmetric discrete light-cone quantization (SDLCQ),
which preserves the supersymmetry at every step of the calculation.
For small and intermediate $r$ the correlator
converges rapidly for all couplings. At small $r$ the correlator
behaves like $1/r^6$, as expected.  At large
$r$ the correlator is dominated by the BPS states of the theory. We find a
critical coupling where the large-$r$ correlator goes to zero;
it grows like the square root of the transverse cutoff.
}

\section{SDLCQ of ${\cal N}=1$ SYM(2+1)}

In this note we display the results of numerical calculations of the 
stress-energy correlator in ${\cal N}=1$ SYM theory in three dimensions.
The motivation for this calculation is twofold. Firstly, correlators of
gauge-invariant operators, such as the stress-energy tensor, are of 
particular interest for string/field theory correspondence scenarios
because they can be evaluated in both of the (conjectured dual) theories
\cite{hlpt00}. 
Secondly, the correlator uses {\em all} spectral information, and therefore
tests also the wave functions of the states, not only the masses.
Additionally, it is technically non-trivial to go in a numerical approach to
higher dimensional problems. 

The method of SDLCQ has been described elsewhere \cite{lup99}, so we can be 
brief here. Following the DLCQ approach \cite{BPP}, we use the 
light-cone coordinates
\begin{eqnarray*}
x^\pm\equiv\frac{1}{\sqrt{2}}(x^0\pm x^1), \qquad x^{\perp}&=& x^{\perp},
\end{eqnarray*}
where $x^+$ plays the role of a time and $x^-$ is a spatial coordinate.
In the sequel, the total longitudinal momentum is denoted by $P^+$, $P^-$ is 
light-cone energy or the Hamiltonian, and $P^\perp$ the
total transverse momentum. 
It was shown in Ref.~\cite{Sakai95} that          
DLCQ preserves supersymmetry, if the supercharge $Q^-$ rather than the
Hamiltonian $P^-$ is discretized. Supersymmetry is the most important 
symmetry in this calculation, because it keeps the theory finite even at 
finite discretization parameter, and we shall follow this approach. 

Let us now formulate ${\cal N}=1$ SYM(2+1) within this framework.
The action is
\[
S=\int d^2 x \int_0^l dx_\perp \mbox{tr}(-\frac{1}{4}F^{\mu\nu}F_{\mu\nu}+
{\rm i}{\bar\Psi}\gamma^\mu D_\mu\Psi),
\]
where the spinor has two components 
$(\psi,\chi)$. We use the light-cone gauge, $A^+=0$. 
We express everything 
in terms of the physical degrees of freedom, namely $\psi$ and 
$\phi\equiv A^\perp$, and can derive the 
light-cone supercharges 
\begin{eqnarray*}\label{sucharge}
Q^+&=&2^{1/4}\int dx^-\int_0^l
dx_\perp\mbox{tr}\left[\phi\partial_-\psi-\psi\partial_-
                 \phi\right],\\
\label{sucharge-}
Q^-&=&2^{3/4}\int dx^-\int_0^l dx_\perp\mbox{tr}\left[2\partial_\perp\phi\psi+
          g_{\rm YM}\left({\rm
i}[\phi,\partial_-\phi]+2\psi\psi\right)\frac{1}{\partial_-}\psi\right]
\end{eqnarray*}
 which fulfill the supersymmetry algebra
\[
\{Q^+,Q^+\}=2\sqrt{2}P^+,\qquad \{Q^-,Q^-\}=2\sqrt{2}P^-, \qquad
\{Q^+,Q^-\}=-4P_\perp.\label{superr}
\]
To evaluate the theory on a computer, we need a finite-dimensional 
Hamiltonian operator. This is achieved by discretizing the theory in 
the following way.
We compactify $x^-$ on a circle of period $2L=2\pi K/P^+$, 
where the harmonic resolution $K$ is effectively a cutoff in particle number.
The longitudinal momenta can therefore take integer values $n_{i}=1,2 ..., K$,
in proper units.
The transverse direction $x^\perp$ is
compactified on circle of period $l$, with the transverse cutoff being
$2\pi T/l$. This cutoff is symmetric, and allows the  
transverse momenta to take values $n^{\perp}_{i}=0,\pm 1,\pm 2, ... ,\pm T$.  
It is a fundamentally different cutoff in the sense that
it is only a momentum, and not a particle-number cutoff.

Close inspection of the resulting supercharges reveals two exact symmetries,
namely transverse parity, which accounts for a 'parity doubling',
and a non-degenerate $S$-Symmetry, associated with a flip of the 
color indices.

\section{Correlation functions}

We want to compute an expression of the form
\begin{equation}\label{T++}
F(x^+,x^-,x^\perp) = \langle 0| T^{++}(x^+,x^-,x^\perp) 
T^{++}(0,0,0)|0 \rangle,
\end{equation}
where $T^{++}(x)$ is a component of the stress-energy tensor, namely the 
momentum operator
\[
T^{++}(x) =  {\rm tr} \left[ (\partial_- \phi)^2 + {1 \over 2} \left(i
\psi \partial_- \psi  - i  (\partial_- \psi) \psi
\right)\right]=T^{++}_B(x)+T^{++}_F(x)\,.
\]
Both the boson and the fermion contributions are two-body operators,
%
%
so that 
only two-particle states will contribute to the correlation function.

To evaluate the correlator, Eq.~(\ref{T++}), we take the collinear 
limit, $x^\perp = 0$, and
insert complete set of states $| \alpha \rangle$ with 
energy eigenvalues $P^-_\alpha$ into the r.h.s.~of  Eq.~(\ref{T++}).
It is straightforward to calculate the correlator, Eq.~(\ref{T++}), in the
free case and it then shows the expected $1/r^6$ behavior for small distances.
The real challenge is the evaluation of the correlator when the complete
set of bound states is inserted in  Eq.~(\ref{T++}).
We now have to write
\begin{eqnarray} \label{eq:Fdiscrete}
F(x^+,x^-,0)&=&\sum_{n,m,s,t} \,\,\left(\frac{\pi}{2L^2 l}\right)^2
\langle 0|\frac{L}{\pi}T^{++}(n,m) e^{-iP^-_{op}x^+-iP^+x^-}
            \frac{L}{\pi} T^{++}(s,t)| 0 \rangle\,,
\end{eqnarray}
where $P^-_{op}$ is Hamiltonian operator. We introduce the convenient notation
\[
\frac{1}{N_u}|u\rangle \propto \frac{L}{\pi}T^{++}(n,m) |0 \rangle.
\]
Inserting a complete set of bound states $|\alpha\rangle $
with masses $M_\alpha$,
and evaluating the sums over $K$ and $N_\perp$ as integrals 
yields
\begin{eqnarray} 
\frac{1}{\sqrt{-i} }\left(\frac{x^-}{x^+} \right)^2 F(x^+,x^-,0)
=\sum_\alpha 
\frac{1}{2 (2\pi)^{5/2}}\frac{M_\alpha^{9/2}}{\sqrt{r}}K_{9/2}(M_\alpha r)
\left[\frac{|\langle u|\alpha\rangle|^2}{lK^3 |N_u|^2}\right]. 
\label{master}
\end{eqnarray}
The term in square brackets is constructed in such a way that it is 
free of any unphysical constants. It is calculated numerically, 
and multiplied by a function including the modified Bessel function 
$K_{9/2}$. Collecting powers of $r$ reveals that the individual 
terms in the sum of Eq.~(\ref{master}) 
behave like $1/r^5$ in the small $r$ limit.
%
\begin{figure}
\centerline{\psfig{file=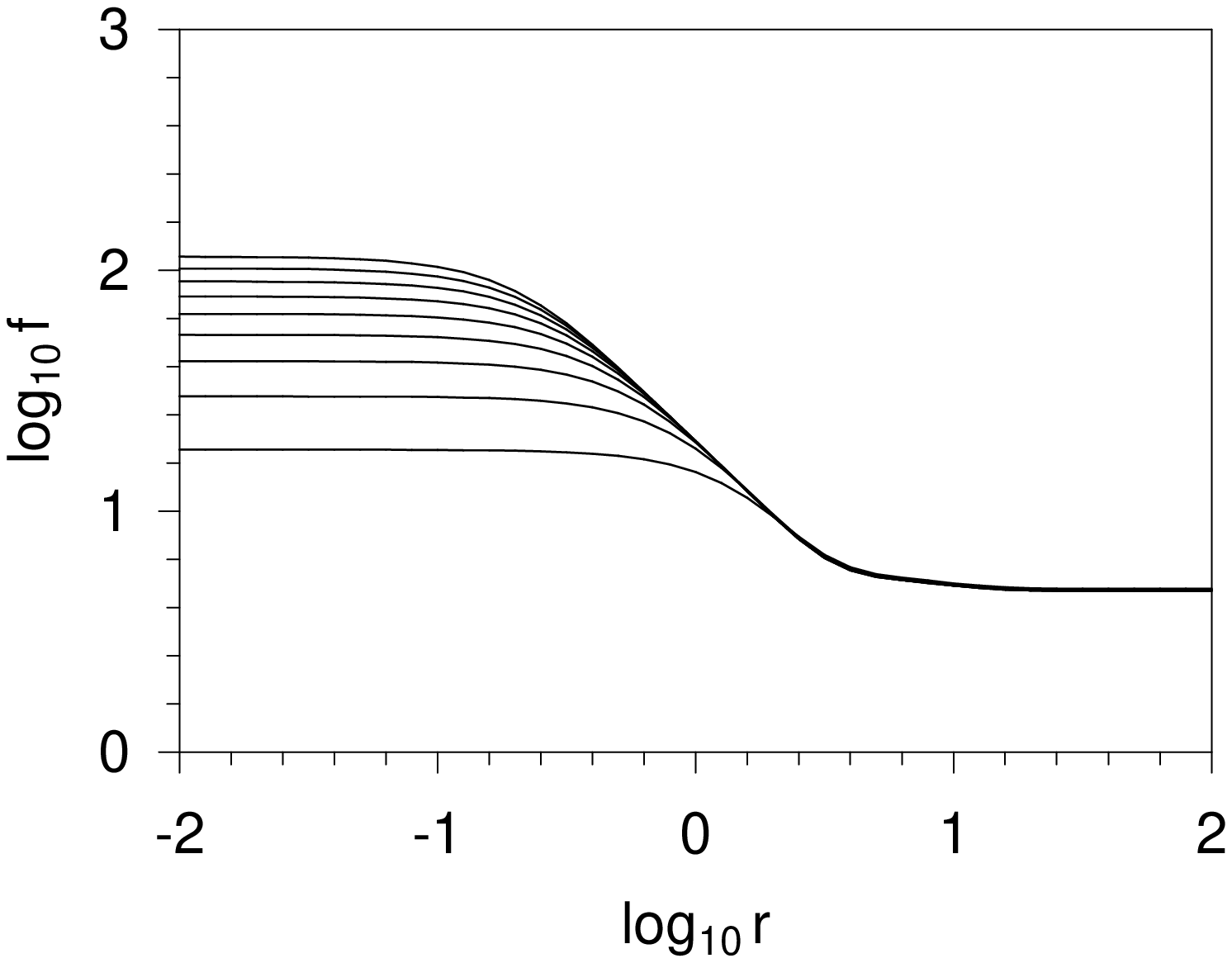,width=2.8in} 
\psfig{file=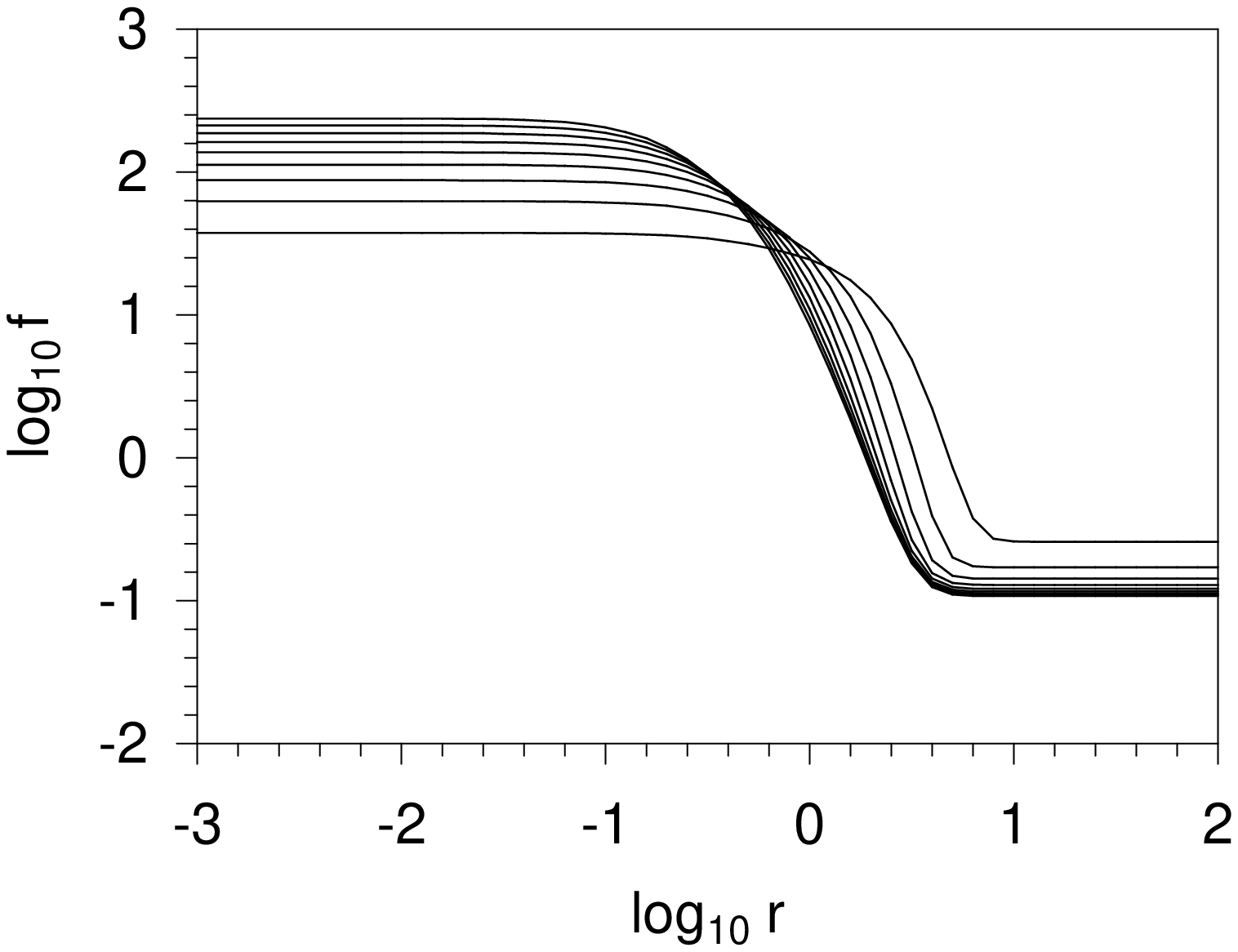,width=2.8in}}
\caption{The log-log
plot of the correlation function $f\equiv r^5\langle T^{++}(x) T^{++}(0) \rangle
\left({x^- \over x^+} \right)^2 \frac{16\pi^3}{105}{{K^3 l} \over \sqrt{-i}}$
vs.\ $r$ (a) for $g{=}0.10$ at $K{=}4$;
(b) for $g{=}1$ at $K=5$. The transverse cutoff runs from $T=1$ to 9.
\label{Fig1}}
\end{figure}
%


\section{Numerical Results}

Let us now look at the numerical results, Fig.~\ref{Fig1}. 
First we focus on the small 
distance behavior of the correlator.
Here, the correlator converges from below with increasing $T$.
We expect a $1/r^6$ behavior from the 
free particle case, but as we saw in the last
section, each individual bound state behaves like $1/r^5$. 
Therefore a coherent behavior of all the states has to take place. This
makes the reproduction of the $1/r^6$ behavior a non-trivial check.
It turns out that the theory passes this test, see Fig.~\ref{Fig1}(a):
for $ -0.5 \leq \log r \leq 0.5 $ the plot of $r^5$ times the
correlator falls like $ 1/r$. 
At larger $K$, Fig.~\ref{Fig1}(b),
we see the same behavior at smaller $r$ as expected. 

Now let us look at the correlation function at large distances.
This region is totally determined by massless states.
There are actually two types of massless states in the theory. 
The massless states at vanishing coupling 
$g=g_{\rm YM} \sqrt{N_c l}/2\pi^{3/2}=0$ are mere 
reflections of the states of the dimensionally
reduced theory. They behave as $g^2 M^2_{1+1}$, where $M^2_{1+1}$
are the masses of the two-dimensional theory \cite{hhlp99}.
We therefore expect for $g \simeq 0$  no
dependence of the correlator on the transverse momentum cutoff
$T$ at large $r$. This is exactly what we see in Fig.~\ref{Fig1}(a).
The other kind of massless states are
exactly massless for all couplings and are BPS states. 
Since they are guaranteed to be massless by the BPS symmetry,
they have to have a complicated dependence on the coupling $g$ through 
their wavefunction, since we see that the large $r$ behavior of the
correlator changes with $g$.
In other words, 
the correlator provides us with information on wave functions of the BPS
states. 

\begin{figure}
\centerline{
\psfig{file=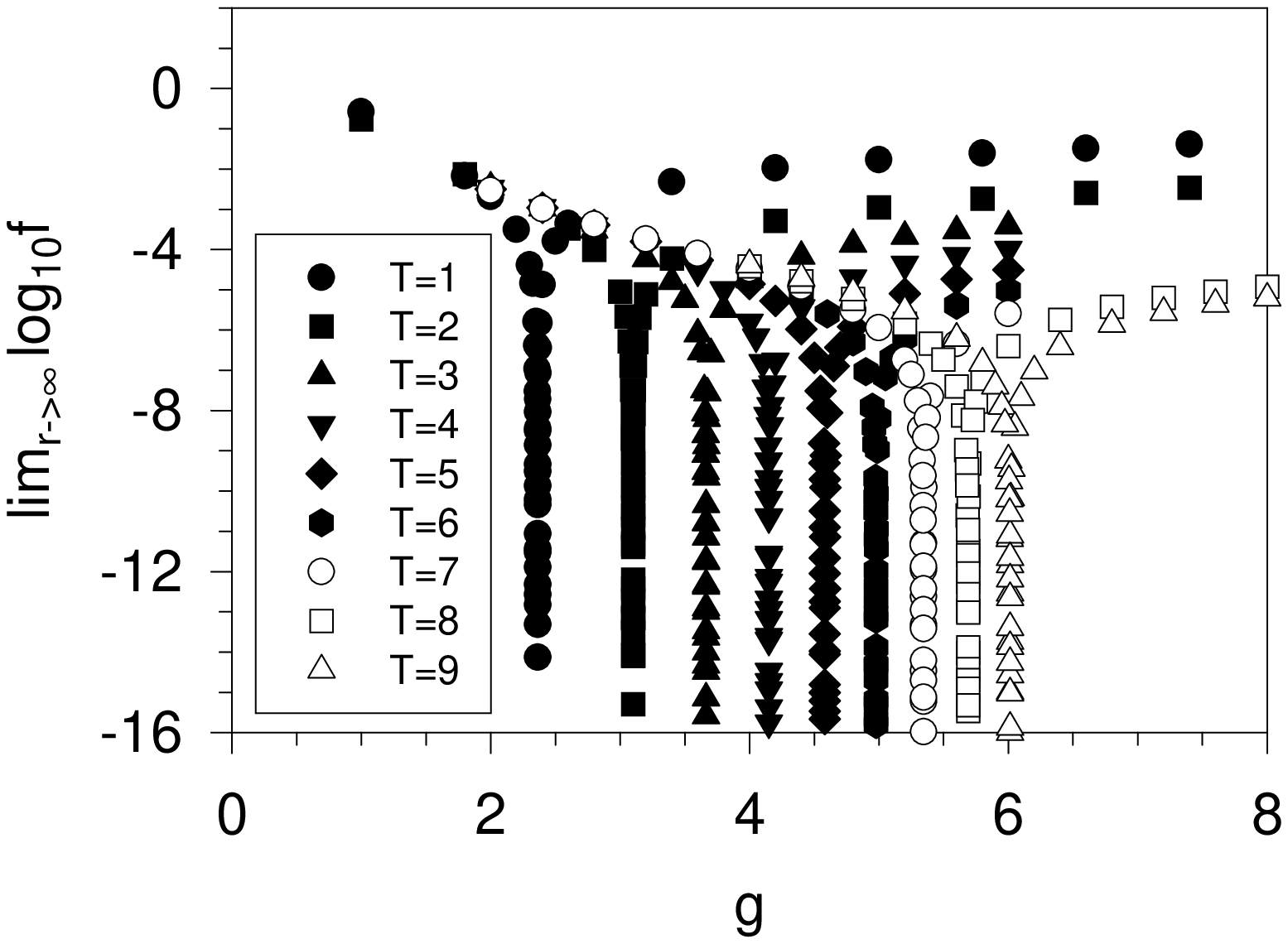,width=2.8in} 
\psfig{file=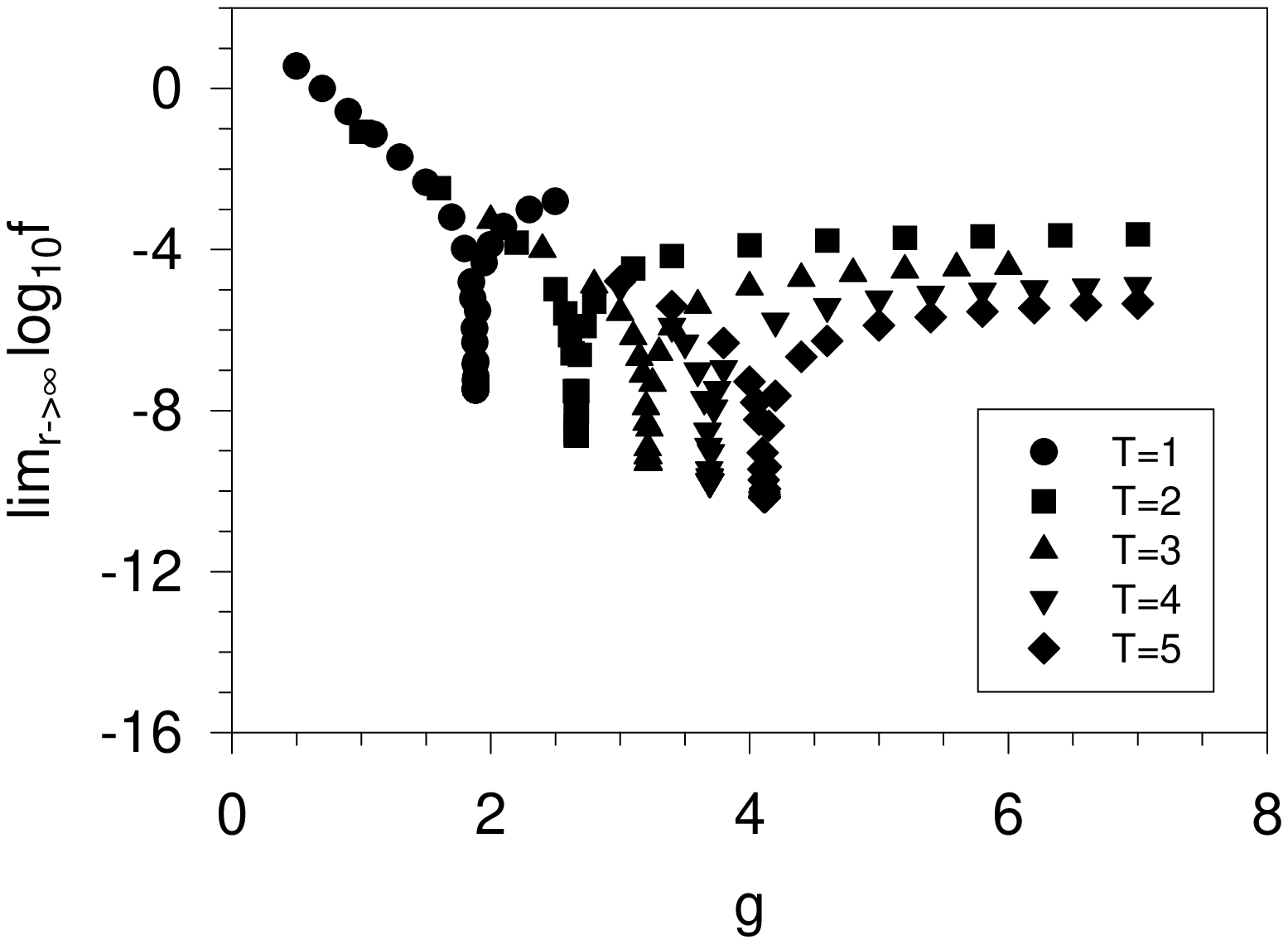,width=2.8in}}
\vspace*{-0.8cm}
\caption{The large-$r$ limit of 
$f\equiv r^5\langle T^{++}(x) T^{++}(0) \rangle
\left({x^- \over x^+} \right)^2 
\frac{16\pi^3}{105}{{K^3 l} \over \sqrt{-i}}$ 
vs.\ $g$ for (a) $K=5$ (b) $K=6$.
\label{Fig2}}
\end{figure}

It turns out that the 
coupling dependence of the large-$r$ limit of the correlator
is very interesting. 
In Fig.~\ref{Fig2} we show the large $r$ limit of the correlator as a function 
of the coupling for two different longitudinal cutoffs.
Surprisingly, the logarithm of the  
correlator does not change monotonically with $g$, but
has singularity at a 'critical' coupling which is a function of $K$ and $T$.
If we plot the `critical' couplings vs.~$\sqrt{T}$  
we find that coupling is a linear function
of $\sqrt{T}$ and the coefficient is largely independent of the cutoff $K$. 


\section{Conclusions}

In this note we showed the first calculation of the stress-energy 
correlator from first principles in three-dimensional ${\cal N}=1$ SYM
theory. 
We recover the $1/r^6$ behavior of the free particle correlator, because
the individual $1/r^5$ behaviors of the bound states add up coherently.
The BPS states show an interesting behavior: although their masses are fixed
by the BPS symmetry, their wavefunctions have to depend on the
coupling, as can be inferred by the correlator data.
Surprisingly, we find a vanishing of the correlator at large distances 
at a critical coupling $g_{crit}\propto \sqrt{T}$.

A word on the computer code may be in order. The latest version of the code
handles two million states, and uses all the present symmetries, 
which in turn 
reduces the size of the matrix to be diagonalized by a factor of eight.
We are therefore optimistic that calculations in full 3+1 dimensions 
are within reach.
Another interesting theory in three dimensions is 
${\cal N}=(8,8)$ SYM(2+1), which is conjectured to correspond
to a string theory of D2-branes, and we are working to generalize
our code to tackle the problems associated with a large number of 'flavors'.  
Our current work should result soon in a complete (low-lying) 
spectrum of ${\cal N}=1$ SYM(2+1),
including all the information in the wavefunctions.
These results could then be compared to future lattice results, namely the 
supersymmetric glueball spectrum. 



\begin{thebibliography}{9}
%
\bibitem{hlpt00} F. Antonuccio, O. Lunin, S. Pinsky, and A. Hashimoto,
JHEP {\bf 07} (1999) 029;
J.R. Hiller, O. Lunin, S. Pinsky, U. Trittmann, 
{\em Phys.\ Lett.}\ {\bf B482} (2000) 409.
\bibitem{Sakai95}
Y. Matsumura, N. Sakai, and T. Sakai,
{\em Phys.\ Rev.}\ {\bf D52} (1995) 2446.
%
%
  {\em Phys.\ Lett.}\ {\bf B429} (1998) 327, hep-th/9803027.
%
%
\bibitem{lup99}
O. Lunin and S. Pinsky,
{\it `` SDLCQ: Supersymmetric Discrete Light Cone Quantization"}
in the proceedings of 11th International Light-Cone School and Workshop
(NuSS 99), Seoul, Korea, 26 May - 26 Jun 1999
(New York, AIP, 1999), p.~140, hep-th/9910222.
%
\bibitem{hhlp99} P. Haney, J.R. Hiller, O. Lunin, S. Pinsky, and U. Trittmann,
{\em Phys. Rev.}\ {\bf D62} (2000) 075002.
%
%
\bibitem{BPP} S.J. Brodsky, H.-C. Pauli, and S.S. Pinsky,
{\em Phys.\ Rep.}\ {\bf301} (1998) 299.
%
%
\end{thebibliography}
\end{document}